\documentclass[10pt,letterpaper]{article}

\usepackage{graphicx}
\usepackage{opex3}

\begin{document}

\title{A high-accuracy algorithm for designing arbitrary holographic atom traps}

\author{Matthew Pasienski and Brian DeMarco}

\address{Department of Physics, University of Illinois at Urbana-Champaign, \\ Urbana, Illinois, 61801}

\email{bdemarco@uiuc.edu}

\begin{abstract}
We report the realization of a new iterative Fourier-transform algorithm for creating holograms that can diffract light into an arbitrary two-dimensional intensity profile.  We show that the predicted intensity distributions are smooth with a fractional error from the target distribution at the percent level.  We demonstrate that this new algorithm outperforms the most frequently used alternatives typically by one and two orders of magnitude in accuracy and roughness, respectively.  The techniques described in this paper outline a path to creating arbitrary holographic atom traps in which the only remaining hurdle is physical implementation.\end{abstract}


\section{Introduction}
The intensity profile of optical fields can be controlled by
using a computer-generated digital phase raster, also called a computer-generated hologram (CGH). The raster, or kinoform, changes the complex phase of the optical field at each pixel without attenuating the amplitude. The CGH, physically implemented using a spatial light modulator (SLM) or similar technology, is illuminated by a monochromatic beam which is relayed to a focusing objective; the desired intensity distribution is produced at the
focal plane of the objective.  CGHs have been used in a wide range of technological applications and physical research. For example, soft matter and biological systems have been manipulated using dynamic arrays of focused beams (i.e., optical tweezers) created using CGHs (see \cite{Grier2003} for a review), and CGHs have been used to shape laser beams for inertial confinement fusion experiments \cite{Dixit1,Lin1, Liu1}.

Although CGH technology has recently been applied to experiments involving ultra-cold atom gases, that work has been limited to arrays of optical dipole traps \cite{Davidson1999,Berg1,Walker2005,Boyer1}.  New applications of optical dipole potentials for cold-atom research, such as interferometers \cite{McGloin1} and atom ``transistors" \cite{Holland2007}, may be realized if high-quality arbitrary light intensity profiles could be generated.  The trapping potential for an atom confined in an optical dipole trap is proportional to the light intensity, and therefore the properties of the intensity profile created by a CGH are critical for such applications.  Desirable features of a CGH applied to trapping ultra-cold atoms include accuracy in matching the desired intensity profile, smoothness of the profile generated by the CGH, and efficiency in diffracting light into the target profile. Smoothness, of particular importance for ultra-cold atom experiments, has received limited attention in previous work on CGHs \cite{Kotlyar1,Senth1}. Disorder related to small-length-scale intensity fluctuations in an optical potential will introduce complications for interferometric applications \cite{Leanhardt2002,Zimmerman2002,Leanhardt2003,Aspect2004} and can greatly affect the study of quantum many-body physics (see \cite{Lewenstein2003}, for example).  Previous measures used to analyze the roughness of intensity profiles created using CGHs could only be applied to uniform distributions; in this manuscript, we introduce a new roughness metric appropriate for continuous profiles. Some constraints related to CGHs may be relaxed for ultra-cold atom applications---a CGH in this context is not required to control the intensity in the entire focal plane, since ultra-cold atom gases are typically confined to a finite region of space.

Calculating a CGH to generate a high-quality arbitrary light intensity distribution is a challenging problem, because a CGH cannot be directly computed, in general, from a desired arbitrary intensity profile.  One technique for calculating a CGH when an exact solution is unknown is to use an iterative Fourier transform algorithm (IFTA), which is computationally efficient compared with other methods, such as a direct binary search \cite{Seld1,Legeard1}.  An IFTA predicts the propagation of a beam through an initial kinoform by fast Fourier transform (FFT), and then successively modifies the kinoform based on a comparison between the predicted and desired focal plane intensities.  The most frequently used IFTAs for calculating CGHs are variants of the Gerchburg-Saxton (GS) and Adaptive-Additive (AA) algorithms \cite{Kotlyar1,Wyrowski2,Soifer1,Ripoll1,spalding2008}.

In this paper, we present a new IFTA that we call the ``mixed-region amplitude freedom" (MRAF) algorithm.  The MRAF algorithm typically improves by one order of magnitude on accuracy and and two orders of magnitude on roughness compared with the GS and AA algorithms for continuous target profiles. To our knowledge, no algorithm for creating CGHs with a comparable level of computational complexity  surpasses the MRAF algorithm in measures of accuracy and roughness.  The MRAF algorithm controls intensity in a bounded two-dimensional subset of the focal plane and achieves accuracy at the percent level at typically the cost of a factor of three in efficiency (compared with the GS and AA algorithms).  Because the MRAF algorithm controls the intensity profile in a single plane, this method can only be applied to creating two-dimensional arbitrary optical traps; confinement to the focal plane will require an additional tightly-focused sheet of light. In section \ref{algorithm} of this manuscript the MRAF algorithm is explained in detail, and in section \ref{results} we report on the algorithm performance for six target intensity profiles.

\section{Algorithm}\label{algorithm}

\begin{figure}
\centering
\includegraphics[width=10cm]{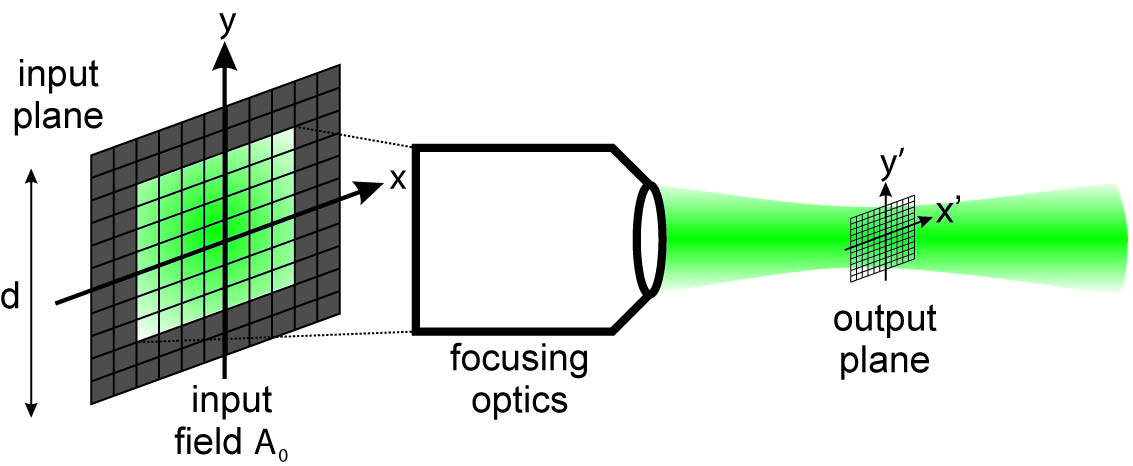}
\caption{Schematic geometry for an IFTA.  The optical field that propagates from the input to the output plane through a focusing objective is shown in green.  The field is discretized using coordinates $(x,y)$ in the input plane and $(x',y')$ in the output plane.  The dashed lines represent the clear aperture of the focusing optics.  The matrix used to computationally represent the input field must be enlarged beyond this region and filled with zero intensity points (dark gray) to fully resolve the output plane.  The physical size of the matrix used to represent the input plane is $d$.} \label{fig1}
\end{figure}

Before giving the mathematical details of the MRAF algorithm, we briefly review the operation of an IFTA \cite{piestun2002}.  An IFTA is a technique to solve the following problem: design a CGH that will convert a light field $A_0(x,y)$ at the CGH, or input plane, into a target intensity distribution $I_0(x',y')$ at the focal, or output, plane of a focusing optic (see Fig.~\ref{fig1})\cite{end1}. The light field $A_0$ is typically a Gaussian beam apodized by the input aperture of the device used to implement the CGH.  The IFTA problem does not have a unique solution, as the complex phase of the optical field associated with $I_0$ is not constrained. This is known as phase freedom; there is a choice of phase in the output plane \cite{Wyrowski3,Wyrowski4}. Complete phase freedom is allowed for far-off resonance optical atom traps because the phase of the field in the output plane does not contribute to the trapping potential if the light is far-detuned from an electronic transition and if the dipole approximation is valid.  An IFTA is designed to use phase freedom to minimize the difference between $I_0$ and the intensity distribution produced by the CGH in the output plane.

\begin{figure}
\centering
\includegraphics[width=10cm]{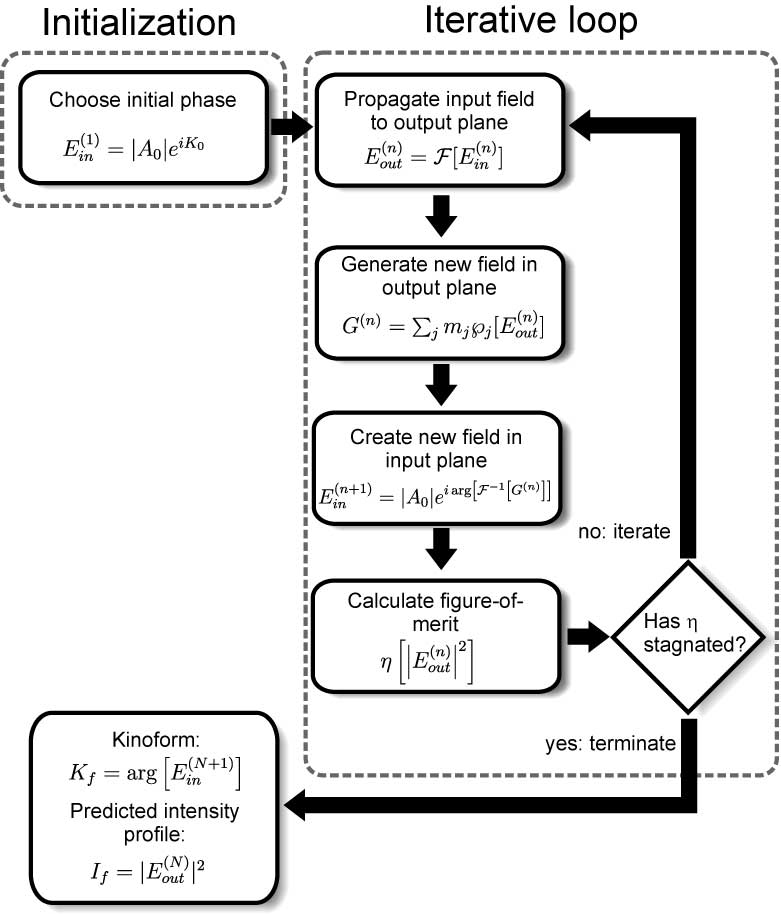}
\caption{Block diagram of an IFTA.} \label{fig2}
\end{figure}

An IFTA can be decomposed into two parts as illustrated in Fig.~\ref{fig2}: an initialization step and an iterative loop. In the initialization step, a phase distribution $K_0(x,y)$ is chosen as a starting point for the algorithm
and is imprinted on $A_0$ to produce the input field $E_{in}^{(1)}(x,y)=|A_0|e^{i K_0}$ for the first iteration. Each iteration $n$ of the loop begins by calculating the field $E_{out}^{(n)}(x',y')=\mathcal{F}\left[E_{in}^{(n)}\right]$ produced by $E_{in}^{(n)}$ propagating to the output plane.  The propagation is modeled using a Fourier transform $\mathcal{F}$, which assumes the paraxial approximation for the focusing optics \cite{Goodman2}.  The algorithm then combines the propagated field $E_{out}^{(n)}$ with the target intensity profile $I_0$ to produce a new field $G^{(n)}(x',y')$.  This procedure is carried out using one or more numerical scalars called mixing parameters $m$.  The phase of the backward propagated field $\arg\left[\mathcal{F}^{-1}\left[G{(n)}\right]\right]$ is used as the starting phase distribution for the next iteration.

The iterative loop is terminated after $N$ iterations once a figure-of-merit $\eta$, calculated using the intensity profile in the output plane and $I_0$, does not improve with repeated iterations.  The phase profile $K_f(x',y')=\arg \left[E_{in}^{(N+1)}\right]$ of the field in the input plane for the final iteration is the kinoform which must be transferred to a physical device.  An ideally implemented CGH will produce the predicted intensity profile $I_f(x',y')=|E_{out}^{(N)}|^2$ in the output plane. The goal of using an IFTA to design a CGH is to choose mixing parameters that optimize one or more measures calculated on the predicted profile $I_f$, such as the deviation from $I_0$.  Achieving this goal will typically require executing an IFTA multiple times with different selections of mixing parameters as part of an optimization scheme \cite{Bigue1}.

Central to the MRAF algorithm is the introduction of amplitude freedom into a restricted region of the output plane and the use of a single mixing parameter. The mathematical details of the MRAF algorithm are given in section \ref{projections}.  The use of amplitude freedom in the MRAF algorithm is not
sufficient to generate a high-quality optical field---a choice of initial phase that eliminates optical vortices from the output plane for all iterations is also necessary.  We outline a procedure for choosing appropriate initial phase distributions in section \ref{phase}.

\subsection{MRAF algorithm}\label{projections}

At each step $n$ of the MRAF algorithm, the propagated field is combined with the target intensity distribution according to:
\begin{equation}
G^{(n)}=\left\{m\sqrt{I_0}|_{SR}+(1-m)\left|E_{out}^{(n)}\right|_{NR}\right\}e^{i\arg\left[E_{out}^{(n)}\right]}.
\end{equation}
A single mixing parameter $m$ controls the relative distribution of optical power in two subsets, the signal region ($SR$) and noise region ($NR$), of the output plane.  Phase freedom is permitted everywhere in the output plane (the phase of the propagated field is used for the phase of $G^{(n)}$), while amplitude freedom is allowed only in the noise region.  Even though the mixing parameter $m$ is kept fixed, the fraction of power in each region changes for every iteration; the only constraints imposed on the MRAF algorithm are that the power in the target profile $\sum{I_0}$ (which is only non-zero in the $SR$) and the total power in the output plane $m\sum_{SR}{\left|E_{out}^{(n)}\right|^2}+(1-m)\sum_{NR}{\left|E_{out}^{(n)}\right|^2}$ remain constant.

The signal region is chosen to overlap with the area in which light will interact with atoms; the remainder of the output plane is the noise region. The effect of dividing the output plane into subsets is to cause the algorithm to converge very closely to the target profile within the signal region, while behaving in a less controlled manner in the noise region. Utilizing the amplitude freedom in the noise region allows for increased accuracy in matching $I_0$ in the signal region, while decreasing the efficiency of the CGH.  The MRAF algorithm is equivalent to a variable strength projection \cite{Buhling2002} or regularized algorithm \cite{Kim2004} with a specific trajectory for the variable mixing parameter that has not been previously demonstrated. The MRAF algorithm is also similar to the algorithm used in \cite{Wyrowski4}, but with a fixed $m$.

In Sec.~\ref{results}, we compare the performance of the MRAF algorithm with the GS and AA algorithms.  The GS algorithm, in which $G^{(n)}=\sqrt{I_0}e^{i\arg\left[E_{out}^{(n)}\right]}$, permits only phase freedom in the output plane.  In the AA algorithm, amplitude freedom is introduced uniformly into the output plane: $G^{(n)}=\left\{m\sqrt{I_0}+(1-m)E_{out}^{(n)}\right\}e^{i\arg\left[E_{out}^{(n)}\right]}$.

\subsection{Initial phase}\label{phase}

As in any optimization scheme, an initial guess that produces a result close to the target improves the convergence rate and reduces the risk of stagnation into a local optimization minimum. Because IFTAs are used when the solution to the CGH problem is unknown, choosing an initial phase profile $K_0$ to reproduce complex features in $I_0$ is not possible. Therefore, we wish to find a $K_0$ as our starting point for which most of the power in the output plane roughly overlaps with the envelope of $I_0$.  The distribution $K_0$ must also be chosen so that $E_{out}^{(1)}$ does not contain any undesired optical vortices---points characterized by a phase singularity and zero intensity---since an IFTA is not able to eliminate vortices present in the output plane \cite{Aagedal1,Kim1,Senth2}.  A further constraint is that only certain choices for $K_0$ can prevent an IFTA from \textit{producing} optical vortices in the output plane at each iteration \cite{Aagedal1,Kim1,Senth2}.  The optical field in an IFTA is discretized and the Fourier transforms are calculated using FFTs.  In order for the FFT to fully resolve the output plane and to reproduce the aperture of the physical device used to implement the CGH, the matrix representing the optical field must be enlarged (i.e., ``padded") beyond the size of the CGH by adding a zero-intensity region \cite{Siegman1975}.  The truncation of the input plane field caused by padding leads IFTAs to create optical vortices in the output plane; this behavior is not completely understood \cite{Aagedal1}.

We find that a quadratic phase distribution for $K_0$ combined with linear and conical gradients does not introduce optical vortices for the MRAF algorithm, even through the input field is truncated in each iteration of the IFTA.  Quadratic phase distributions, the equivalent of a thin lens, were first discussed as a solution to the vortex problem in the context of variants of the GS algorithm \cite{Aagedal1}.  A quadratic phase profile, given by $K_0(x,y)=4R \left[\alpha x^2+\left(1-\alpha\right) y^2\right]$ ($R$ is the curvature and $\alpha/(1-\alpha)$ is the aspect ratio) changes the size of the envelope of the intensity profile in the output plane.  A linear gradient phase profile $K_0(x,y)=B\left[x\cos(\mu)+y\sin(\mu)\right]$, where $B$ is the strength of the gradient and $\mu$ is an angle, shifts the centroid of the intensity profile in the output plane.  A conical phase gradient $K_0(x,y)=Br$ creates a ring in the output plane, where $r=\sqrt{x^2+y^2}$.

For the work in this paper, quadratic phase profiles are used to roughly match the size of the field in the output plane to the size of the target profile.  We find that the results of the algorithm are not strongly affected by small changes in $R$.  Linear gradients are used to match targets that are shifted from the center of the output plane to avoid complications caused by undiffracted light resulting from the finite efficiency of a physically-implemented CGH.  Conical gradients are used to match target profiles which have a ring-like structure.  Quadratic, linear, and conical phase distributions are added together modulo $2\pi$ to combine the effects of each.  We find that this combination of initial phase distributions has enough flexibility to obtain sufficient overlap of $|E_{out}^{(1)}|^2$ with $I_0$ to achieve a few percent error in $I_f$ within tens of iterations.

\section{Results}\label{results}

\begin{figure}
\centering
\includegraphics[width=10cm]{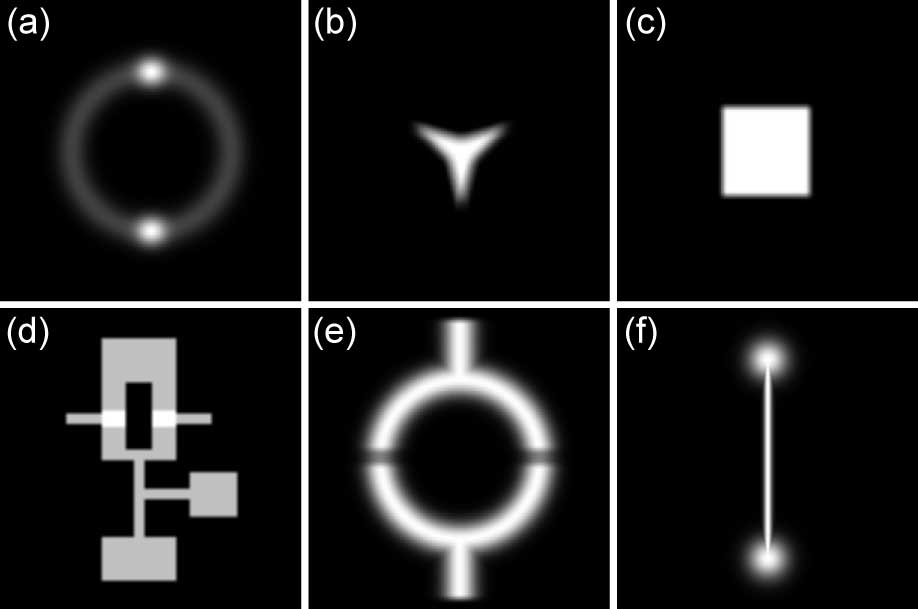}
\caption{Target intensity profiles $I_0$ used to characterize the performance of the MRAF algorithm.  The field-of-view for images (a), (b), (c), and (e) is a $200\times200$ pixel and for (d) and (f) is a $400\times400$ pixel subset of the output plane centered on $I_0$.  The grayscale represents intensity, with black corresponding to the regions of zero intensity. The radius of the ring in (a) is $53$ pixels and the waist for each Gaussian beam and the Gaussian cross-section of the ring is $14$ pixels.  The maximum intensity of the Gaussian beams is three times that of the ring.  Each ``tip" of the star-shaped pattern in (b) is 20 pixels from the center of the star; the two lines that intersect to form each ``tip" subtend a $28^\circ$ angle.  To create the profile shown in (b), a uniform intensity profile with a star shape was convolved with a Gaussian with a 5 pixel waist.  The profile in (c) was created by convolving a 58 pixel on edge square profile with a 3 pixel averaging filter.  The overall dimensions of the profile in (d) are 288 pixels wide and 325 pixels high, and the intensity in the ``base" regions is increased by 33\%.  The Gaussian ring in (e) has a 53 pixel radius and a 7 pixel r.m.s. width.  The intensity in the 10 pixel wide gaps in (e) is suppressed by a factor of 2, and the ``leads" in (e) are 185 pixels from end-to-end.  The 264 pixel wide Gaussian ``wire" in (f) has a 3.5 pixel r.m.s. width, and the Gaussian reservoirs in (f) have a 17.6 pixels r.m.s. radius. The center of each profile is displaced from the center of the output plane by (b) 37 pixels and (c) 63 pixels; the profiles in (a), (d), (e), and (f) are centered on the output plane.} \label{fig3}
\end{figure}

We characterize the performance of the MRAF algorithm using the six target intensity profiles shown in Fig.~\ref{fig3}, chosen because of their potential application to ultra-cold atom experiments. Target (a) consists of two Gaussian beams connected by a ring with a Gaussian cross-section, which could be used to study ultra-cold atom gases in multiply-connected geometries.  A 3-pointed star-shaped intensity profile is shown in Fig.~\ref{fig3}(b); a similar profile was recently used to induce spontaneous vortex generation in an atomic Bose-Einstein condensate (BEC) \cite{Scherer1}.  Fig.~\ref{fig3}(c) shows a uniform square intensity profile, which may be used as an optical lattice beam in experiments for which it is desirable to remove effects generated by the external confinement resulting from a Gaussian beam profile \cite{Jaksch1998,DeMarco2005}.  Target (d) is a complex intensity profile designed to be evocative of an optical lattice beam that could be employed to realize an ``atomtronic" logical OR gate \cite{Holland2007}.  An intensity profile that could be used to trap a BEC in a superconducting quantum interference device (SQUID) geometry is shown in Fig.~\ref{fig3}(e).  Finally, a BEC confined in a dipole trap created using target profile (f) would be equivalent to a thin superconducting wire connected between bulk superconductors \cite{bezryadin00}.

The parameters defining the geometry of each target profile are given in the caption to Fig.~\ref{fig3}.  To describe these test patterns and in the rest of this manuscript, we use pixels ($px$) to measure distances in the input and output planes.  Each pixel in the input plane represents a point at which the input field is discretized.  The physical size of a pixel in the input plane is the physical size, $d$, of the matrix used to represent the input field divided by the number of pixels.  The pixel size in the output plane is $f\lambda/d$, where $f$ is the focal length of the focusing objective and $\lambda$ is the wavelength of the light. For the results given in this section, the input field is discretized on a $768\times768$ pixel array and the phase of the field in the input plane is discretized in 256 levels (for each iteration of the IFTA) \cite{end2}.  The input field array is padded with zero intensity points in each iteration of the IFTA to create a $1536\times1536$ pixel matrix.  We observe no significant change in the algorithm results if the array is enlarged beyond $1536\times1536$ pixels (consistent with the Nyquist-Shannon sampling theorem).  We implement IFTAs in MATLAB \cite{code}, and we use a Gaussian input field $A_0\propto e^{-r^2/w_0^2}$ with a 565 pixel waist $w_0$.

\begin{figure}
\centering
\includegraphics[width=10cm]{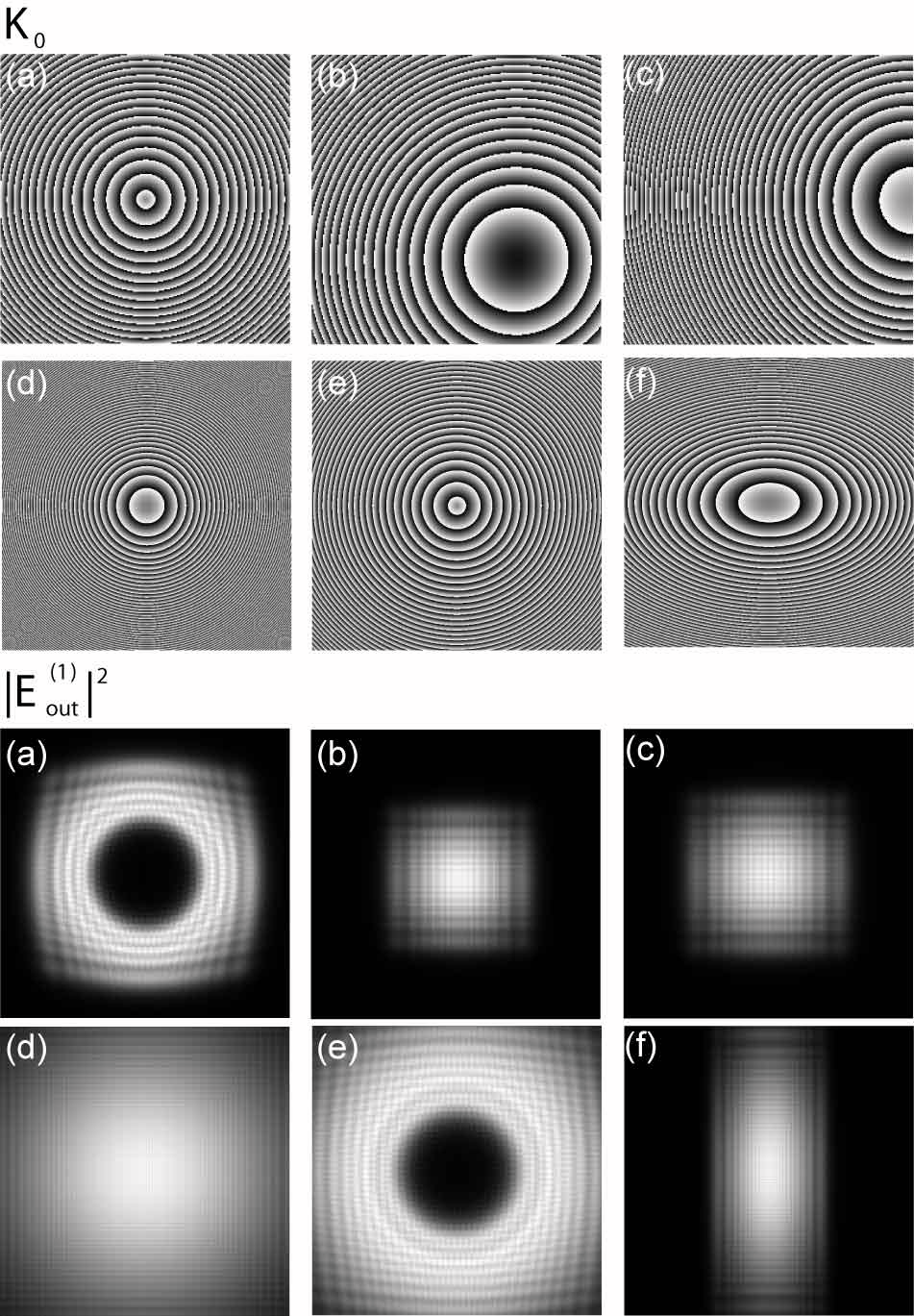}
\caption{Initial phase profiles $K_0$ (top row) and predicted initial intensity profiles $|E_{out}^{(1)}|^2$ (bottom row) chosen for the target intensity profiles in Fig.~\ref{fig3}. The phase profiles are 768$\times$768 pixels and are shown in grayscale modulo $2\pi$; white corresponds to a $2\pi$ phase. Conical phase profiles with $B=117$~$mrad/px$ are employed in (a) and (e).  Linear gradients of 136~$mrad/px$ and 260~$mrad/px$ with $\mu=0$ and $\pi/4$ are used for (b) and (c), respectively.  Quadratic phase profiles with $R=0.31~mrad/px^2$, $R=0.3~mrad/px^2$, $R=0.34~mrad/px^2$, $R=1.4~mrad/px^2$, $R=0.5~mrad/px^2$, and $R=1.6~mrad/px^2$;$\alpha=0.29$ are applied in (a), (b),(c), (d), (e), and (f) respectively.} \label{fig4}
\end{figure}

The initial phase profile chosen for each target intensity profile is shown in the top half of Fig.~\ref{fig4}. A conical phase profile is used in $K_0$ for (a) and (e) to create a ring structure.  Target profiles (b) and (c) are shifted from the center of the output field, and therefore linear gradients are used in $K_0$ to displace the intensity in $E_{out}^{(1)}$ accordingly.  Quadratic phase profiles are used in each $K_0$ to match the approximate area covered by $I_0$ in the output plane.  The intensity profile in the output plane for the first iteration of the IFTA is shown in the bottom half of Fig.~\ref{fig4}. The initial phase profiles in Fig.~\ref{fig4} were optimized manually; the predicted intensity profile is not affected by small changes in $K_0$.

\begin{figure}
\centering
\includegraphics[width=14cm]{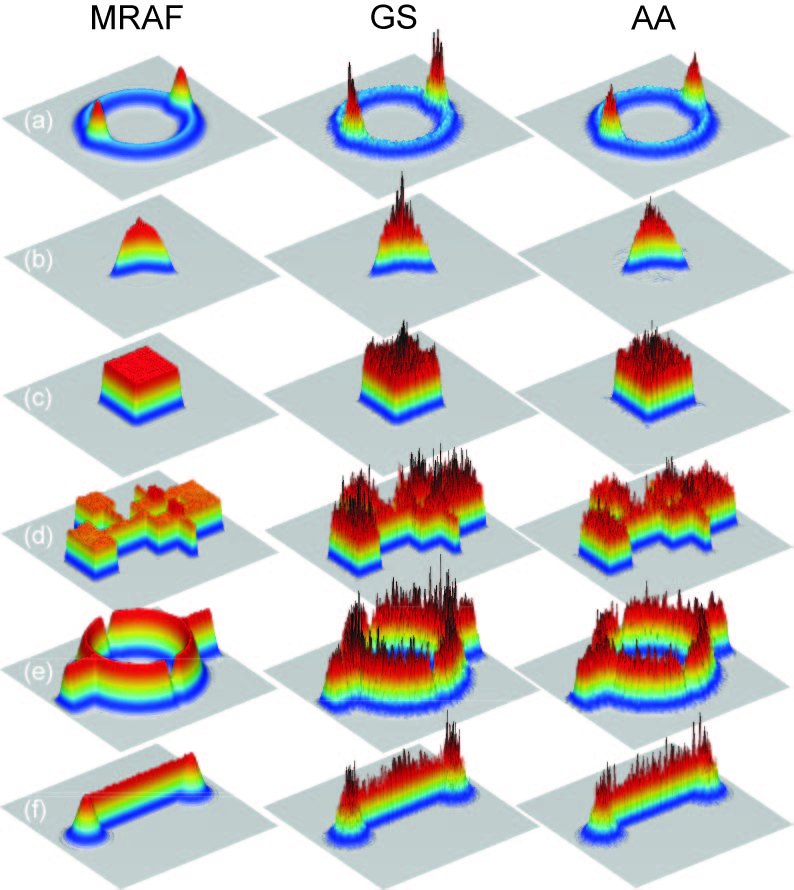}
\caption{Intensity profiles $I_f$ for the MRAF, GS, and AA (left, middle, right) algorithms for the test target profiles.  Only the intensity in the signal region is shown, and the profiles are scaled so that the total power in the signal region is the same for each.} \label{fig5}
\end{figure}

A qualitative comparison between the predicted intensity profiles for the MRAF, GS, and AA algorithms is shown in Fig.~\ref{fig5}.  We find that the small-length-scale intensity fluctuations apparent in Fig.~\ref{fig5} are a generic feature of using the GS and AA algorithms to generate arbitrary intensity profiles. We did not determine if any of these fluctuations are optical vortices, which can be removed from $I_f$ under limited circumstances by changing $K_f$ (and at the cost of greatly increased computational complexity) \cite{Ripoll1,Aagedal1,Senth2,Bertaux1}.

To quantitatively characterize the performance of the MRAF algorithm and compare with the GS and AA algorithms, we introduce measures of accuracy and roughness.  The accuracy metric is the root-mean-square (r.m.s.) fractional error from the target averaged across a subset, the measure region ($MR$), of the output plane:
\begin{equation}
\eta=\sqrt{\frac{1}{N_{MR}}\sum_{(x',y')\in MR}  \frac{\left[\widetilde{I_f}(x',y')-\widetilde{I_0}(x',y')\right]^2}{\widetilde{I_0}(x',y')^2}}.
\label{eta}\end{equation}
The measure region is a subset of the signal region and is chosen to exclude the zero-intensity pixels in $I_0$. The intensity profiles $\widetilde{I_f}=I_f/\sum_{(x',y')\in MR}I_f$ and $\widetilde{I_0}=I_0/\sum_{(x',y')\in MR}I_0$ are normalized to have the same power in the measure region, and $N_{MR}=\sum_{(x',y')\in MR}1$ is the number of pixels in the measure region. The error $\eta$ is also used as the convergence parameter and the optimization parameter for the MRAF and AA algorithms.

The roughness measure
\begin{equation}
\rho=\sum_{(x',y')\in MR}\left\{H\left[\widetilde{I_f}(x',y')-\widetilde{I_0}(x',y')\right]\right\}^2/N_{MR}
\label{S}\end{equation}
is the average of the square of the mean curvature $H$ of the difference between the predicted and target intensity profiles in the measure region.  The roughness $\rho$ is proportional to the Willmore bending energy for the surface $\widetilde{I_0}-\widetilde{I_f}$ \cite{willmore}.  A unique measure of roughness for a two-dimensional manifold, such as $I_f$, does not exist.  We choose $\rho$ as defined in Eq.~\ref{S} as a metric because it is intuitively appealing as an energy that is strongly weighted by small-length scale deviations of $I_f$ from $I_0$ (the Willmore bending energy of a spherical surface is proportional to the square of the inverse of the radius of curvature). The measure $\rho$ also quantitatively reproduces qualitative features we observe in the predicted intensity profiles.  For each predicted intensity profile we also calculate the efficiency of the CGH for diffracting light into signal region.  The efficiency $\xi=\sum_{(x',y')\in SR}\widetilde{I_f}/\sum_{(x',y')} \widetilde{I_f}$ is defined as the ratio of the power in the signal region to the total power in the output plane. The parameters $\eta$ and $\rho$ are both efficiency-independent measures of the deviation of $I_f$ from $I_0$: $\eta$ and $\rho$ are zero if $\widetilde{I_f}=\widetilde{I_0}$.

\begin{figure}[htpb]
\centering
\includegraphics[width=10cm]{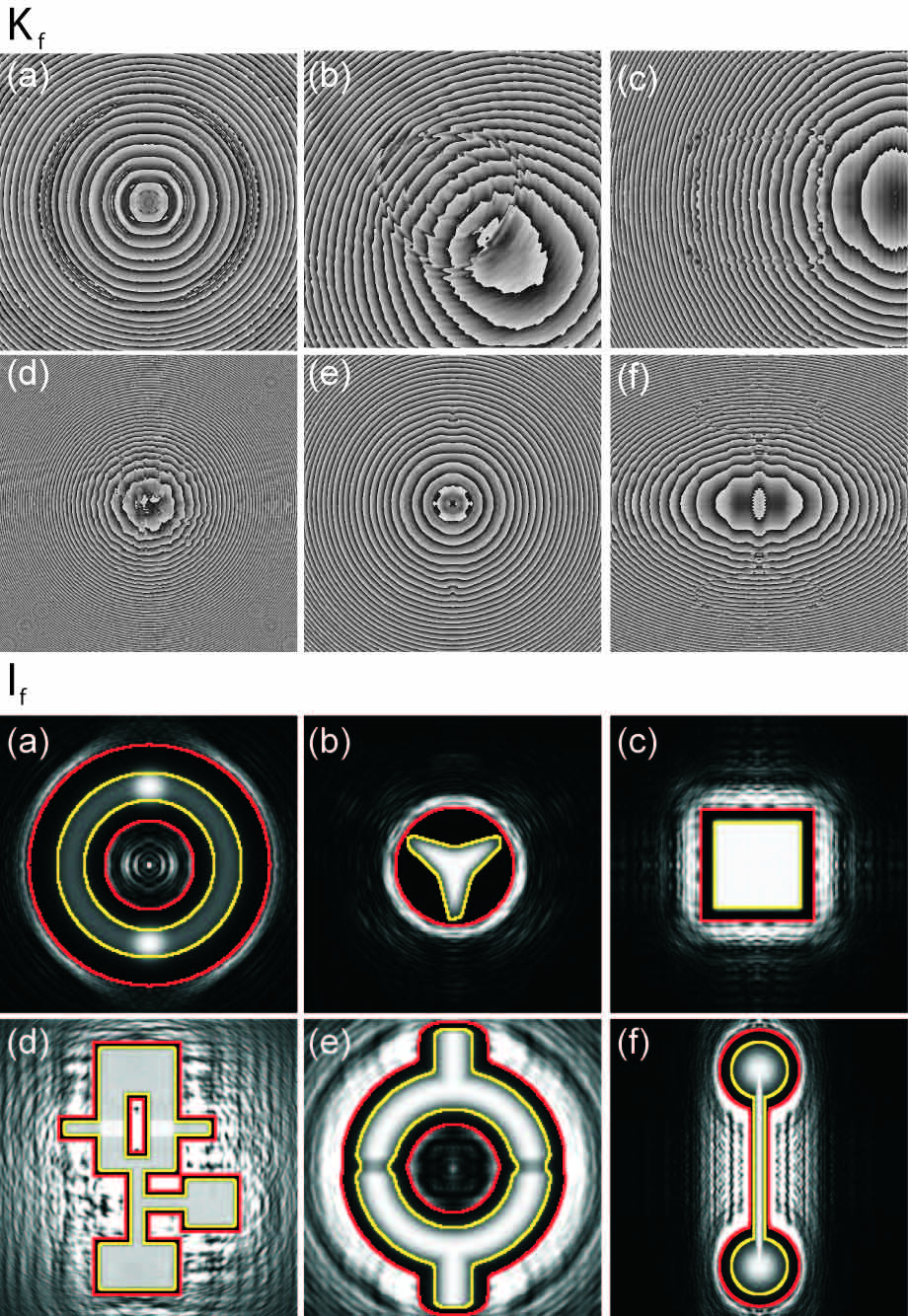}
\caption{Final kinoforms $K_f$ (top row) and predicted intensity profiles $I_f$ (bottom row) produced by the MRAF algorithm for targets (a), (b), and (c).  The mixing parameters used to generate these results are: (a) 0.40, (b) 0.35, (c) 0.40, (d) 0.30, (e) 0.35, and (f) 0.30.  The signal region (red) for (a) is an annulus with with inner and outer radii 25 and 81 pixels; in (b) is a circle with a 40 pixel radius; for (c) is a square 75 pixels on edge; in (d) is a region 10 pixels from the edge of the target profile; for (e) a region 10 pixels away from where the intensity is 10\% of the maximum intensity; and for (f) consists of two 53 pixel radius circles separated by 264 pixels and a connective region 25 pixels wide.  The measure region (yellow) in (a) is an annulus with inner and outer radii 44 and 62 pixels; in (b), (e), and (f) is defined by a region in which the intensity of the target is greater than 10\% of the maximum target intensity; in (c) is a square 57 pixels on edge; and for (d) is the edge of the target profile.} \label{fig6}
\end{figure}

The result of using the MRAF algorithm to create a CGH based on the test target profiles is shown in Fig.~\ref{fig6}.  Both the final kinoforms to which the the initial phase profiles in Fig.~\ref{fig4} converged and the predicted intensity profiles are shown in the figure.  The mixing parameters and the conical and quadratic phase profiles used in $K_0$ were optimized by determining the minimum value for $\eta$ calculated for a wide range of $m$, $B$, and $R$.  The signal region used in the MRAF algorithm is outlined in red for each target profile in Fig.~\ref{fig6}.  The MRAF algorithm converged in less than 100 iterations for each of these target profiles.

\begin{figure}
\centering
\includegraphics[width=10cm]{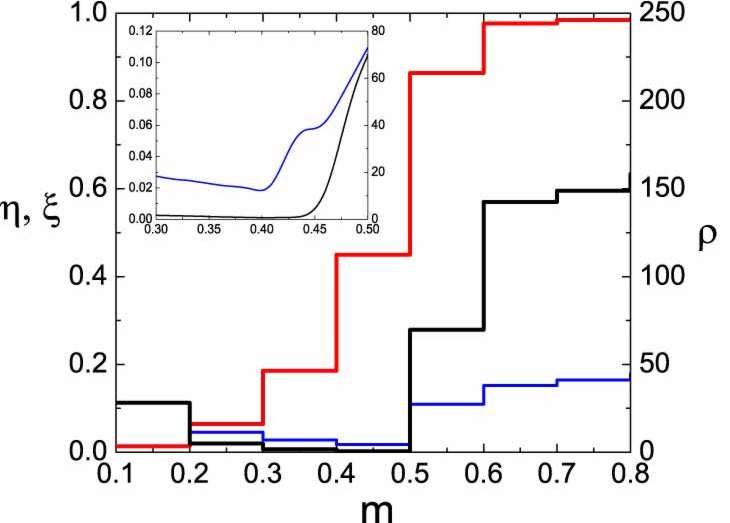}
\caption{Variation of measures characterizing the MRAF algorithm performance as the mixing parameter $m$ is varied.  The efficiency $\xi$ (red), roughness $\rho$ (black), and error $\eta$ (blue) are shown for target (a) for different values of the mixing parameter $m$.  The inset shows detail around the globally-optimized value of $m$. The mixing parameter that minimizes $\eta$ approximately coincides with an minimum in roughness $\rho$ for the MRAF algorithm.} \label{figmix}
\end{figure}

Table 1 shows the accuracy, roughness, and efficiency calculated for each predicted intensity profiles in Fig.~\ref{fig6} and the equivalent results generated by the GS and AA algorithms.  The accuracy and roughness are determined using the measure region outlined in yellow in Fig.~\ref{fig6}.  The MRAF algorithm on average shows a factor of 9 improvement in accuracy and a factor of 190 improvement in roughness compared with the GS and AA algorithms; the average error for each target for the MRAF algorithm is at the few percent level.  The MRAF algorithm produces comparatively smooth intensity profiles even though $\rho$ is not used to optimize the mixing parameters or as a convergence criteria for the IFTA (see Fig.\ \ref{figmix}).

\begin{table}[htpb]
\begin{center}
\begin{tabular}{|l|l|l|l|l|l|l|l|}
\hline
Algorithm & \multicolumn{6}{c|}{Error $\eta$}\\
\cline{2-7}
& Ring (a) & Star (b) & Square (c) & OR gate (d) & SQUID (e) & thin wire (f)\\
\hline
GS & 0.21 & 0.30  & 0.23 & 0.34 & 0.36 & 0.36\\
AA & 0.13 & 0.19  & 0.19 & 0.23 & 0.21 & 0.23\\
MRAF & 0.017 & 0.027  & 0.015 & 0.039 & 0.018 & 0.029\\
\hline

 & \multicolumn{6}{c|}{Roughness $\rho$}\\
\cline{2-7}

GS & 220 & 5600 & 460 & 13 & 160 & 110\\
AA & 64 & 2200 & 400 & 5.3 & 47 & 40 \\
MRAF & 0.65 & 20 & 1.1 & 0.044 & 0.18 & 0.19\\

\hline
 & \multicolumn{6}{c|}{Efficiency $\xi$}\\
\cline{2-7}
GS & 0.99 & 0.99 &  0.99 & 0.96 & 0.97 & 0.97\\
AA & 0.90 & 0.83 & 0.88 & 0.79 & 0.71 & 0.59\\
MRAF & 0.45 & 0.29 & 0.45 & 0.18 & 0.30 & 0.19 \\
\hline
\end{tabular}
\end{center}
\caption{Table comparing the performance of the MRAF to the GS and AA algorithms.  The mixing parameters used for the AA algorithm are (a) 1.9, (b) 2.0, (c) 1.9, (d) 2.0, (e) 2.2, and (f) 2.5.  The GS and AA algorithms converged in 100 iterations for the results in this table.}
\end{table}

\begin{figure}
\centering
\includegraphics[width=12cm]{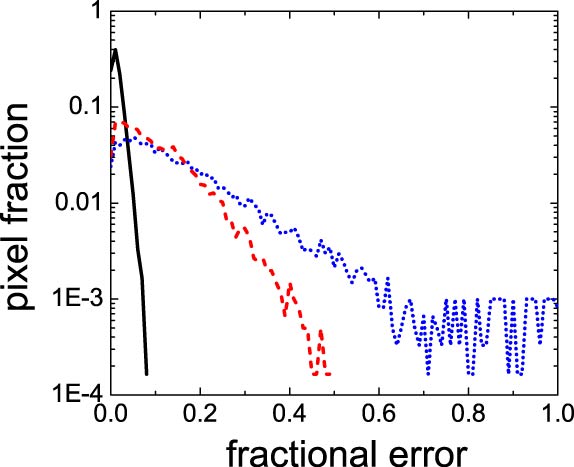}
\caption{Histogram of the fractional error at each pixel evaluated for $I_f$ for the MRAF algorithm used on target (a).  The fraction of pixels in the signal region are binned with respect to the fractional error $\sqrt{\widetilde{I_f}^2-\widetilde{I_0}^2}/\widetilde{I_0}$.  The width of each bin is equivalent to a 1\% fractional error.  The solid black, blue dotted, and red dashed lines are the result for the MRAF, GS, and AA algorithms, respectively.} \label{fig7}
\end{figure}

The error $\eta$, an average across the measure region, can be small although large errors exist at points in the output plane. To show that the MRAF algorithm achieves accuracy everywhere in the signal region, a histogram of the error evaluated at each output plane point ($\sqrt{\widetilde{I_f}^2-\widetilde{I_0}^2}/\widetilde{I_0}$) in the signal region for the result in Fig.~\ref{fig6}(a) is shown in Fig.~\ref{fig7}; 95\% of the pixels in the signal region have less than a 3\% error.  For the purposes of comparison, the result of using the GS and AA algorithms to calculate a CGH for target (a) is also shown in Fig.~\ref{fig6}.  The MRAF algorithm improves greatly on the GS and AA algorithms, both of which have at least 45\% of the pixels in the signal region with errors greater than 10\%.

For the targets used in this manuscript, the MRAF algorithm has approximately a factor of 2--3 lower efficiency compared with the GS and AA algorithms.  While the MRAF algorithm does not lead to efficiencies as high as the GS and AA algorithms, it does not perform so poorly that the efficiency advantage of using a MRAF is lost.  For example, the authors of \cite{Scherer1} created a profile similar to (b) by propagating a Gaussian beam through an intensity mask and imaging the result onto a BEC.  The efficiency of that mask in transmitting light was 3\%, compared with $\approx29$\% for a CGH created using the MRAF algorithm.

\section{Conclusion}

In conclusion, we have reported the realization of a new IFTA for designing CGHs that can be used to create arbitrarily-shaped, two-dimensional optical dipole traps for ultra-cold atom experiments.  The MRAF algorithm has relatively low computational complexity and converges rapidly---within tens of iterations.  For six test target profiles, the predicted output of a CGH designed using the MRAF algorithm is comparatively smooth and has errors at the percent level.

The challenge for realizing arbitrary, two-dimensional dipole traps for atoms now lies with experimentally implementing a CGH designed using the MRAF algorithm. In this paper, we assume abberation-free optics, the paraxial approximation, a single-polarization optical field, and ideal CGH response. Some or all of these idealizations will be violated in an experimental realization of a CGH, leading to the intensity profile produced at the focal plane deviating from the predicted profile $I_f$.  The extent to which these practical considerations affect an experimental implementation will depend on the details of the kinoform and the specific nature of technical problems.

In particular, non-ideal CGH response is likely to have a high-impact on the quality of the intensity profile produced in an experiment. In this paper, we take into account two practical limitations of CGH technology: quantized phase levels and finite resolution.  Many approaches to producing CGHs also do not have well-characterized or well-controlled phase response. For example, commercially-available, scientific-grade SLMs are afflicted by non-uniform and nonlinear phase response \cite{Inoue2006}.  Controlling these problems at the percent level to experimentally achieve the high-accuracy of the MRAF algorithm will be challenging.

\section*{Acknowledgments}

We thank Paul Kwiat, Benjamin Lev, David McKay, and Matthew White for critically reading this manuscript.  This work was supported by the Office of Naval Research and the National Science Foundation (Grant No. 0448354). Any opinions, findings, and conclusions or recommendations expressed in this material are those of the authors and do not necessarily reflect the views of the National Science Foundation.


\begin{thebibliography}{10}
\bibitem{end1}
We introduce several quantities in this manuscript which are functions of the
  input and output plane coordinates. For notational simplicty, we indicate
  this functional dependence only for the first use and when quantitites are
  explicitly written as functions of the coordinates.
\bibitem{end2}
A $768\times768$ pixel kinoform and 256 discrete phase levels were used to
  match a commercially available, scientific-grade SLM: the Hamamatsu X8267.
\end{thebibliography}
\end{document}